\documentclass[runningheads]{llncs}
\usepackage{graphicx}

\usepackage{pgfplots}
\usepackage{tikz}
\usepackage{mathrsfs}
\usepackage{amsfonts}
\usepackage[ruled]{algorithm2e}
\usepackage{floatrow}
\newfloatcommand{capbtabbox}{table}[][\FBwidth]

\SetCommentSty{mycommfont}
\renewcommand{\thefootnote}{}
\begin{document}
	%
	\title{Multi-Modality Pathology Segmentation Framework: Application to Cardiac Magnetic Resonance Images}
	%
		\author{Zhen Zhang \inst{1} \and
		Chenyu Liu \inst{1} \and
		Wangbin Ding \inst{1} \and
		Sihan Wang \inst{2} \and
		Chenhao Pei \inst{1} \and
		Mingjing Yang \inst{1,*} \and
		Liqin Huang\inst{1}}
	
	
	\institute{College of Physics and Information Engineering, Fuzhou University, Fuzhou, China 
		\and School of Basic Medical Science, Fudan University, Shanghai, China
	}
	\footnotetext{* M Yang is the corresponding authors: yangmj5@fzu.edu.cn}
	%
	%
	%
	%
	\maketitle              
	\begin{abstract}
      Multi-sequence of cardiac magnetic resonance (CMR) images can provide complementary information for myocardial pathology (scar and edema). However, it is still challenging to fuse these underlying information for pathology segmentation effectively. This work presents an automatic cascade pathology segmentation framework based on multi-modality CMR images. It mainly consists of two neural networks: an anatomical structure segmentation network (ASSN) and a pathological region segmentation network (PRSN). Specifically, the ASSN aims to segment the anatomical structure where the pathology may exist, and it can provide a spatial prior for the pathological region segmentation. In addition, we integrate a denoising auto-encoder (DAE) into the ASSN to generate segmentation results with plausible shapes. The PRSN is designed to segment pathological region based on the result of ASSN, in which a fusion block based on channel attention is proposed to 
     better aggregate multi-modality information from multi-modality CMR images. Experiments from the MyoPS2020 challenge dataset show that our framework can achieve promising performance for myocardial scar and edema segmentation.
    		
	\end{abstract}
	\renewcommand{\thefootnote}{\arabic{footnote}}
	\setcounter{footnote}{0}
	\section{Introduction}
	
	
	Myocardial infarction (MI) is one of the most dangerous cardiovascular diseases in worldwide. The severity of MI depends on the assessment of the myocardial scar and edema \cite{berry2010magnetic}. Accurate delineation of these pathological regions from cardiac magnetic resonance (CMR) can provide important advancements for the prediction and management of MI patients \cite{ingkanisorn2004gadolinium}. Since manual delineation is generally time-consuming, tedious and subject to inter-observer variations, the automatic segmentation approach has gradually attracted more attention of research.
	
	Conventional myocardial pathology segmentation methods are mainly based on intensity thresholding, such as the signal threshold to reference mean (STRM) \cite{kolipaka2005segmentation}, region growing (RG) \cite{alba2012healthy} and full-width at half-maximum (FWHM) \cite{amado2004accurate}. However, the thresholding methods could be easily affected by the image noise, and have poor agreement with expert delineations \cite{spiewak2010comparison,zhang2016myocardial}.
	Recently, learning-based methods have achieved promising performance in different pathology segmentation tasks, such as brain tumor \cite{wang2017automatic} and liver lesion \cite{zeng2019liver}. For pathology segmentation on left atrium (LA) myocardium, Yang et al. presented a super-pixel scar segmentation method using support vector machine (SVM) \cite{yang2018fully}. Li et al. proposed a fully automated scar segmentation method based on the graph-cuts framework, where the potentials of the graph are estimated via deep neural network (DNN) \cite{li2020atrial}.
	Futhermore, Li et al. designed a multi-task learning network to joint perform LA segmentation and LA scars quantification, in which the LA boundary is extracted as spatial attention for the scars \cite{li2020joint}. 
	For pathology segmentation on left ventricular (LV) myocardium, Zabihollahy et al. proposed a  CNN-based method to segment scar from late gadolinium enhancement (LGE) MRIs \cite{zabihollahy2019convolutional}. However, their method relies on the manual delineation of the LV myocardium region. To achieve a fully automatic scar segmentation method, they further developed a multi-planar network to segment LV myocardium \cite{zabihollahy2020fully}. 

\begin{figure}[htb] 
		\centering
		\includegraphics[width=0.94\textwidth]{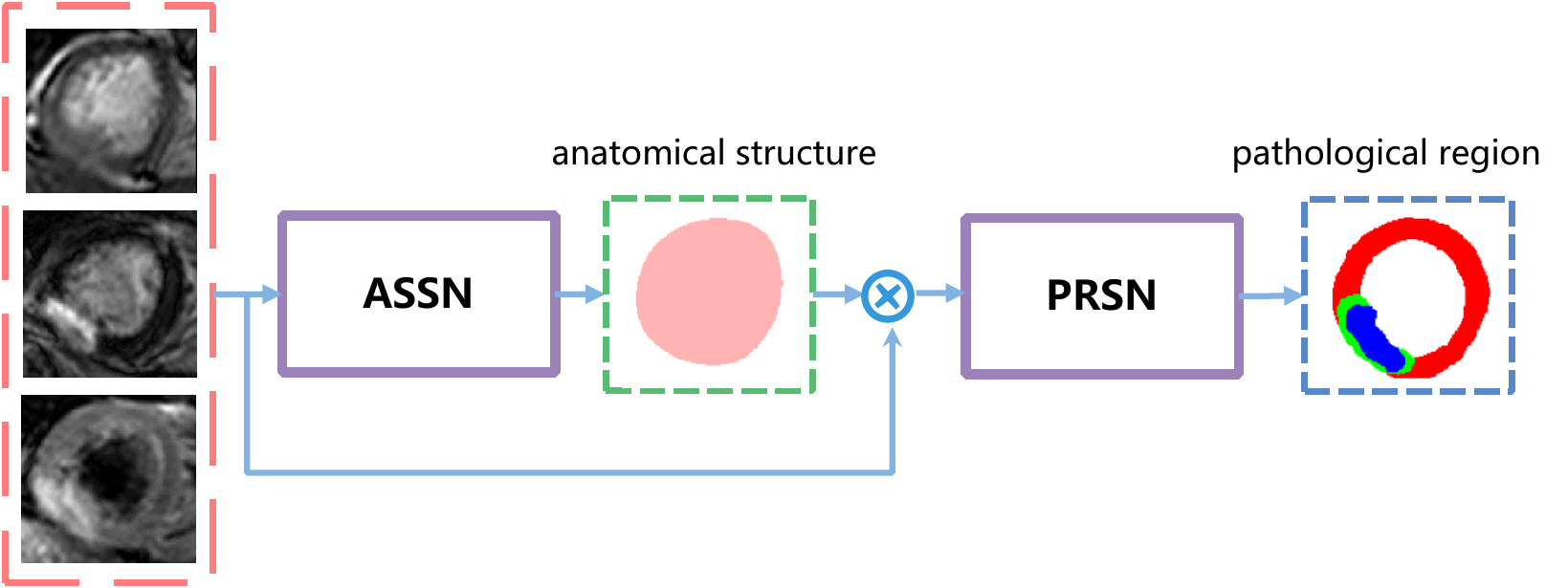}
		\caption{The architecture of the multi-modality pathology segmentation framework. Given multi-modality CMR (bSSFP, T2, DE) images, the ASSN first obtains a candidate anatomical structure, where the pathology may exist. Then, the PRSN predicts the final scar and edema regions within the candidate structure.}
		\label{fig:1}    
	\end{figure}

	
At present, most DNN-based myocardial pathology segmentation methods are focus on mono-modality CMR, such as LGE. But multi-modality CMR can provide different enhanced-information of the whole heart. For instance, the balanced-Steady State Free Precession (bSSFP) cine sequence can present a clear myocardial boundary, while the LGE and T2-weighted CMR can highlight the scar and edema regions, respectively \cite{zhuang2016multivariate,zhuang2018multivariate}. Being aware that the complementary information is helpful for myocardial pathology segmentation. We design a cascade multi-modality pathology segmentation framework. Figure \ref{fig:1} shows the overview of the framework. The framework decomposes the pathology segmentation task into two sub-stages, i.e. the anatomical structure segmentation stage and the pathological region segmentation stage.  The main contributions of our work are: 

(1) we propose a fully automatic pathology segmentation framework, and validate it using the MyoPS2020 challenge dataset \footnote{http://www.sdspeople.fudan.edu.cn/zhuangxiahai/0/MyoPS20/}. 

	(2) we present an anatomical structure segmentation network, where a denoising auto-encoder (DAE) is adopted to reconstruct the segmentation results with realistic shapes. 
	
	(3) we propose a pathological region segmentation network, in which a channel attention based fusion block is designed to adaptively fuse complementary information of multi-modality CMR images for pathology segmentation.
	
	\section{Method}
	\subsection{Anatomical Structure Segmentation Network (ASSN)}
	\label{sec:reg}
	
\begin{figure}[htb] 
		\centering
		\includegraphics[width=1\textwidth]{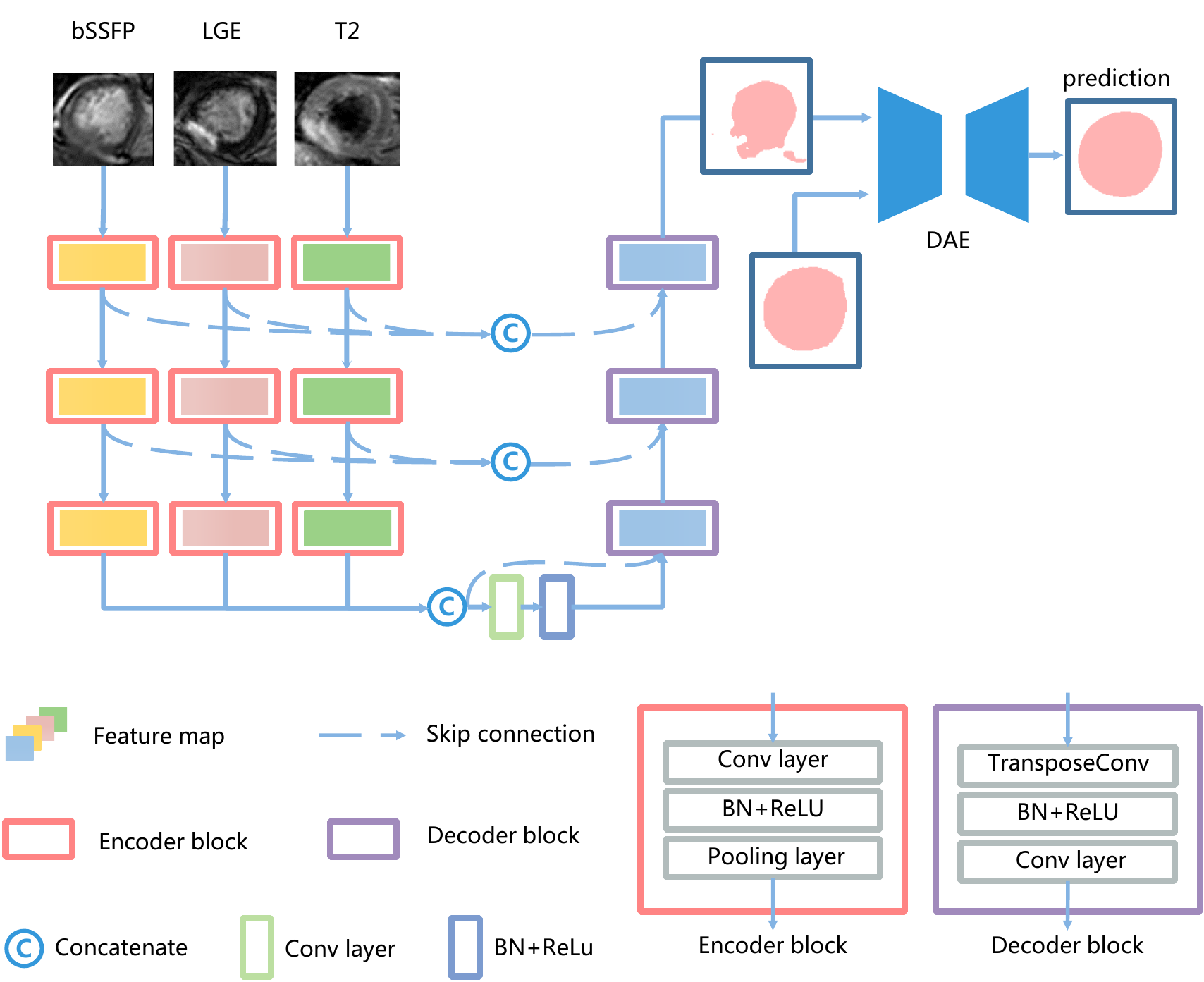}
		\caption{The architecture of the ASSN. The auxiliary DAE is adopted to suppress the influence of the pathology region and generate results with plausible shape.}
		\label{fig:2}    
	\end{figure}

	The ASSN is designed to obtain a candidate anatomical structure from CMR images. In the myocardial pathology segmentation task, we designate the candidate structure as the LV epicardial region, where the scar and edema may exist. Figure \ref{fig:2} shows the architecture of the ASSN. It mainly includes three individual encoders and one shared decoder. Each encoder can obtain underlying anatomical feature from CMR, while the decoder can fuse the obtained features, and predicts a pixel-level LV mask.
	
	Given a multi-modality CMR images $I$=($I_{bSSFP}$, $I_{LGE}$, $I_{T2}$), the ASSN aims to learn a mapping $f_\theta$ from $I$ to a binary mask. Therefore, the network can be trained under supervised manner, and the loss function is
	\begin{equation}
		\mathcal{L}oss_{seg}=\mathcal{D}ice(f_\theta(I),L_{lv})
	\end{equation}
	where the $L_{lv}$ is the golden standard of the LV, $\mathcal{D}ice(A,B)$ refers to the Dice score of $A$ and $B$. Thus, the candidate anatomical structure $C$=($C_{bSSPF}$, $C_{LGE}$, $C_{T2}$) can be extracted as 
	\begin{equation}
		C_{bSSFP}=I_{bSSFP}\otimes f_\theta(I),
	\end{equation}
		\begin{equation}
C_{LGE}=I_{LEG}\otimes f_\theta(I),
	\end{equation}
		\begin{equation}
C_{T2}=I_{T2}\otimes f_\theta(I),
	\end{equation}
	where $\otimes$ is element-wise multiplication. 
	
	Generally, the ASSN performs pixel-wise classification based on processing the intensity value of $I$. However, the pathology usually leads to abnormal intensity distribution in CMR images. For instance, LGE visualizes the scars as brighter texture, in contrast to the dark healthy myocardium \cite{zhuang2020cardiac}. Therefore, the segmentation results could be easily affected. To tackle this, we adopt a DAE to refine the segmentation results with realistic shapes \cite{yue2019cardiac}. 
    
	A DAE usually follows an encoder-decoder (E-D) architecture. Let the $\ddot{L}$ denotes the noisy version of the $L_{gd}$, the DAE aims to map the $\ddot{L}$ to a lower-dimension representation $h$, from which the $L_{gd}$ can be reconstructed. It can be trained to minimize the reconstruction error of the input
	\begin{equation}
	\label{equ:dae}
		\mathcal{L}oss_{DAE}=\|D(E(\ddot{L}))-L_{gd} \|_2,
	\end{equation}
	where $E(\ddot{L})$ is a compact representation of $\ddot{L}$, $D(E(\ddot{L}))$ is a reconstruction of $L_{gd}$. 
	Regarding the original segmentation result $f_\theta (I)$ as a noisy version of the golden standard label, we integrate the DAE into the ASSN to reconstruct the original result into a plausible one. So that, the final loss function of the ASSN is defined as
	\begin{equation}
	\label{equ:assn}
		\mathcal{L}oss_{ASSN}=\mathcal{L}oss_{seg}+ \beta\mathcal{D}ice(D(E(f_\theta (I))),L_{lv}),
	\end{equation}
	where $ \beta$ is the balance coefficient between the Dice loss and reconstruction loss. 
		
	\subsection{Pathological Region Segmentation Network (PRSN)}
	\label{sec:sim}
\begin{figure}[tb] 
	\centering
	\includegraphics[width=1.0\textwidth]{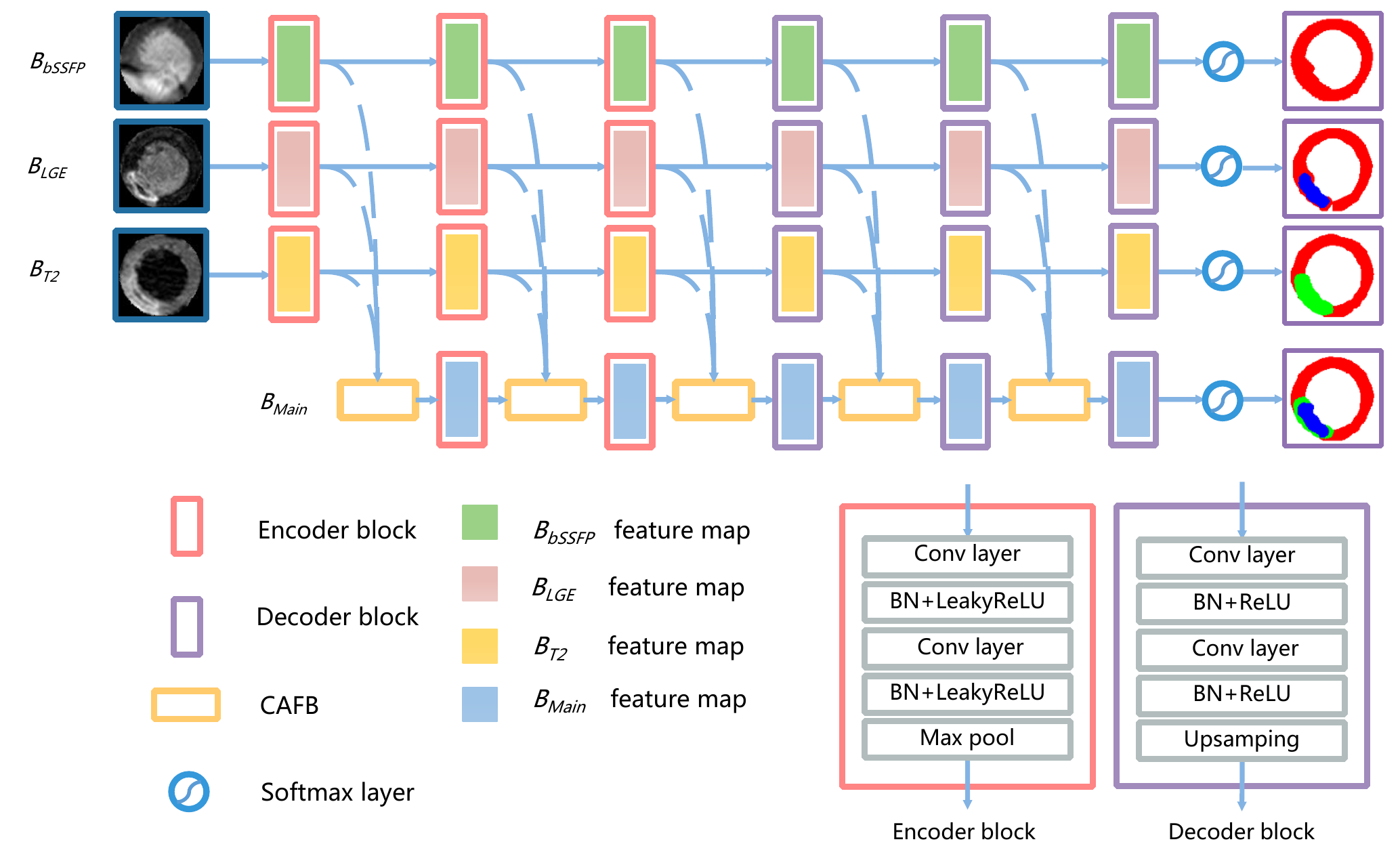}
	\caption{The architecture of PRSN. It contains four sub-branches $B_{bSSFP}$, $B_{LEG}$, $B_{T2}$, $B_{Main}$.  We set up to segment the myocardium in $B_{bSSFP}$; the scars and normal myocardium in $B_{LEG}$; the edema and normal myocardium in $B_{T2}$; the scars, edema and normal myocardium in $B_{Main}$.}
		\label{fig:3}    
	\end{figure}

	In the pathological region segmentation network (PRSN), complementary information from
	$C=(C_{bSSPF},C_{LGE},C_{T2})$ are expected to be fused and boost the pathology segmentation performance. 
	Figure \ref{fig:3} shows the architecture of PRSN. We construct three DNN branches $(B_{bSSPF},B_{LGE},B_{T2})$ to capture multi-modality information from each candidate region. Specifically, the $B_{LEG}$ and $B_{T2}$ mainly aim to acquire pathology (scar and edema) features from $C_{LGE}$ and $C_{T2}$. Meanwhile, the $B_{bSSFP}$ is intent to obtain myocardium features from $C_{bSSPF}$. Due to most of the scar and edema are scatted in the myocardium, the myocardium features can provide spatial prior information for the pathology regions. Therefore, the cost functions of three DNN branches are
		\begin{equation}
	    \mathcal{L}oss_{bSSFP}=\mathcal{D}ice(L_{bSSFP},\hat{L}_{bSSFP}),
		\end{equation}
			\begin{equation}
		\mathcal{L}oss_{LGE}=\mathcal{D}ice(L_{LGE},\hat{L}_{LGE}),
		\end{equation}
			\begin{equation}
		\mathcal{L}oss_{T2}=\mathcal{D}ice(L_{T2},\hat{L}_{T2}),
		\end{equation}
	where $L_{bSSFP}$ ($L_{LGE}$, $L_{T2}$) and $\hat{L}_{bSSFP}$ ($\hat{L}_{LGE}$, $\hat{L}_{T2}$ ) are corresponding gold standard and predicted label of $B_{bSSPF}$ ($B_{LEG}$, $B_{T2}$) branch, respectively.

\begin{figure}[tb] 
	\centering
	\includegraphics[width=0.9\textwidth]{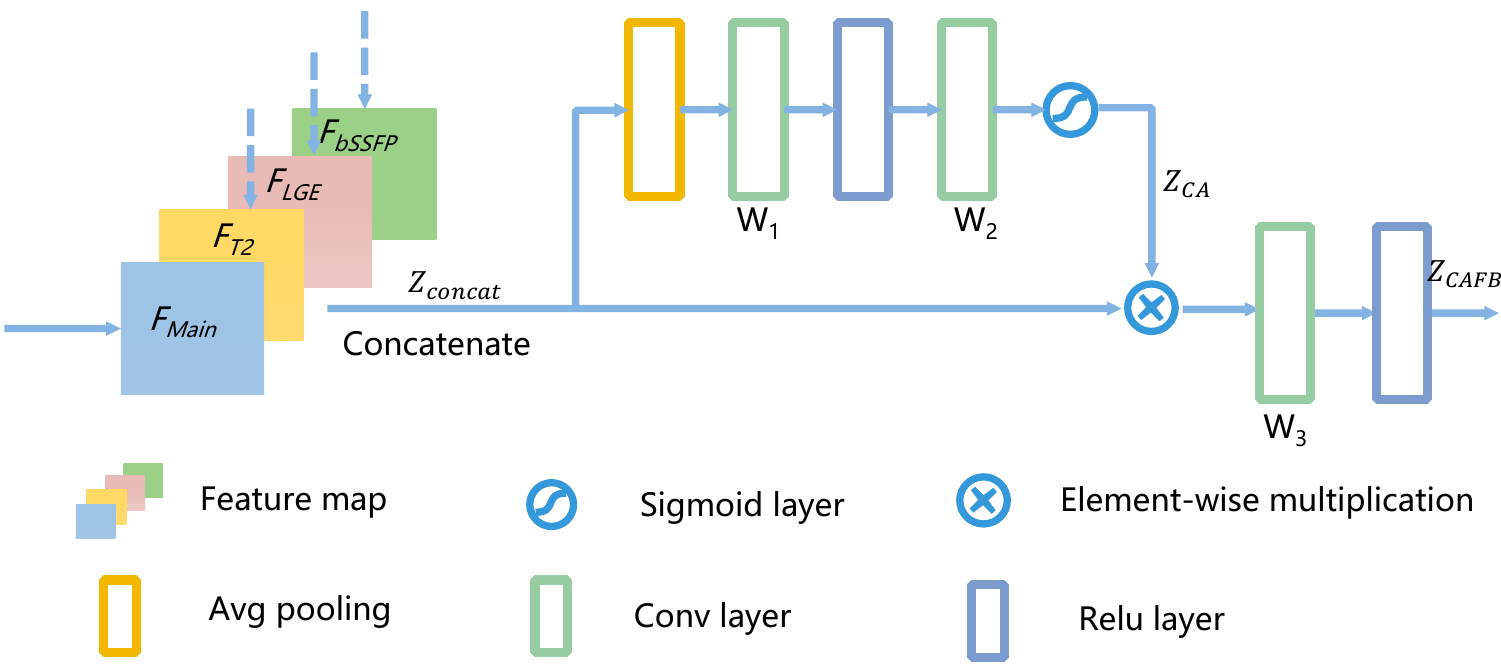}
	\caption{The channel-wise fusion block}
	\label{fig:cafm}    
\end{figure}

Having the three sub-branches constructed, the potential features from them are need be fused and propagated to a main-branch $B_{Main}$ for pathology segmenting. At present, the most popular feature fusion strategies include summation, product and maximization. However, they still suffer from the lack of robustness in different tasks \cite{zhou2020hi}. As shown in the Figure \ref{fig:cafm}, we propose a multi-modality channel-attention fusion block (CAFB) for adaptively weighted feature fusion of different modalities in $B_{Main}$. 

Suppose we have three feature maps $(F_{bSSFP}, F_{LGE}, F_{T2})$ from sub-branches and one previous output $ F_{Main}$ of the main-branch, the CAFB first merges these feature maps to obtain an concatenated feature $Z_{concat}$. Since $Z_{concat}$ aggregates all feature maps from $(B_{bSSFP}$, $B_{LEG}$, $B_{T2})$, it easily suffers from the information redundancy. Due to the channel-attention (CA)\cite{hu2018squeeze} can emphasize informative features and suppress less useful ones, the block adopts it to performs channel-wise feature re-calibration on $Z_{concate}$. So That, the output of CAFB ($Z_{CAFB}$) is
	\begin{equation}
		Z_{CA}= \delta(W_2 \sigma(W_1 Avg(Z_{concat})),
	\end{equation}
	\begin{equation}
		Z_{CAFB}= \sigma(W_3 (Z_{CA}\otimes Z_{concat})),	
	\end{equation}
where $\sigma$, $\delta$ and $Avg$ refer to Relu, Sigmoid and average pooling function, respectively; and $W_1$, $W_2$ and $W_3$ are parameters of different convolution layers.
Here, $Z_{CA}$ is the channel-wise attention weight, with which the original $Z_{concat}$ can be re-calibrated and achieve better representation of multi-modality information.

Furthermore, we apply the CAFB in different hierarchies of the $B_{Main}$ (see $B_{Main}$ in Figure \ref{fig:3}). Thus, the $B_{Main}$ can capture multi-scale multi-modality features for pathology segmentation. The training loss of the  $B_{Main}$  can be defined as 
\begin{equation}
	\mathcal{L}oss_{Main}=\mathcal{D}ice(L_{Main},\hat{L}_{Main}),
\end{equation}
where $L_{Main}$ and $\hat{L}_{Main}$ are the gold standard and predicted label of $B_{Main}$, respectively. Note that the $B_{Main}$ jointly performs scar, edema and normal myocardium segmentation. Finally, the overall loss function of the PRSN is 
\begin{equation}
    \label{equ:prsn}
	\mathcal{L}oss_{PRSN}=\mathcal{L}oss_{Main}+\lambda_{bSSFP}\mathcal{L}oss_{bSSFP}+\lambda_{LGE}\mathcal{L}oss_{LEG}+\lambda_{T2}\mathcal{L}oss_{T2},
\end{equation}
where $\lambda_{bSSFP}$, $\lambda_{LGE}$ and $\lambda_{T2}$ are hyper-parameters.

	\section{Experiment}
	\subsection{Dataset}
The framework was evaluated in the MyoPS2020 challenge data set which contains 45 multi-sequence CMR (bSSFP, LGE, T2) images. In our experiments, the training and validation sets  comprise 20 and 5 images with their gold standard labels, respectively. The testing set includes 20 images. All published data has been aligned in a common space and resampled with the same spatial resolution.
For evaluation, we randomly selected 20 samples for training networks, while leaving the rest of 5 samples for validation.
	\subsection{Implementations}
	We trained our models by extracting 2D slices from multi-sequence CMR images. Each slice was cropped and resized to $128\times 128$ pixels which are roughly centering at the heart region. All of the models (DAE, ASSN and PRSN) were implemented in Python and optimized by using the Adam algorithm.
	
	For the DAE: In each training iteration, we generated $\ddot{L}$ by randomly adding noise to a gold standard label $L$. Having a pair of ($L$, $\ddot{L}$) prepared, the DAE can be trained via minimizing the reconstruction loss $\mathcal{L}oss_{DAE}$ (see Eq.\ref{equ:dae}).
	
	For the ASSN: The pre-trained DAE was adopted to perform shape reconstruction. In each training iteration, the sample $I=(I_{bSSFP},I_{LGE},I_{T2},L)$ was feed into the network. By setting the $\beta$ to $0.2$ in $\mathcal{L}_{ASSN}$ (see Eq.\ref{equ:assn}), the trainable loss can be calculated and back-propagated to optimize the parameters of ASSN. 

	For the PRSN:  We first extracted $C=(C_{bSSPF},C_{LGE},C_{T2})$ and their corresponding label $(L_{bSSPF},L_{LGE},L_{T2})$ from the training data. Then, the hyper-parameter $\lambda_{bSSFP}$, $\lambda_{LGE}$ and $\lambda_{T2}$ were set to  $0.3$, $0.5$ and $0.5$, respectively (see Eq. \ref{equ:prsn}). Finally, the network can be trained by minimizing the  $\mathcal{L}oss_{PRSN}$.

\subsection{Results}
\subsubsection{ASSN:}
\begin{figure}[htb] 
		\centering
		\includegraphics[width=1.0\textwidth]{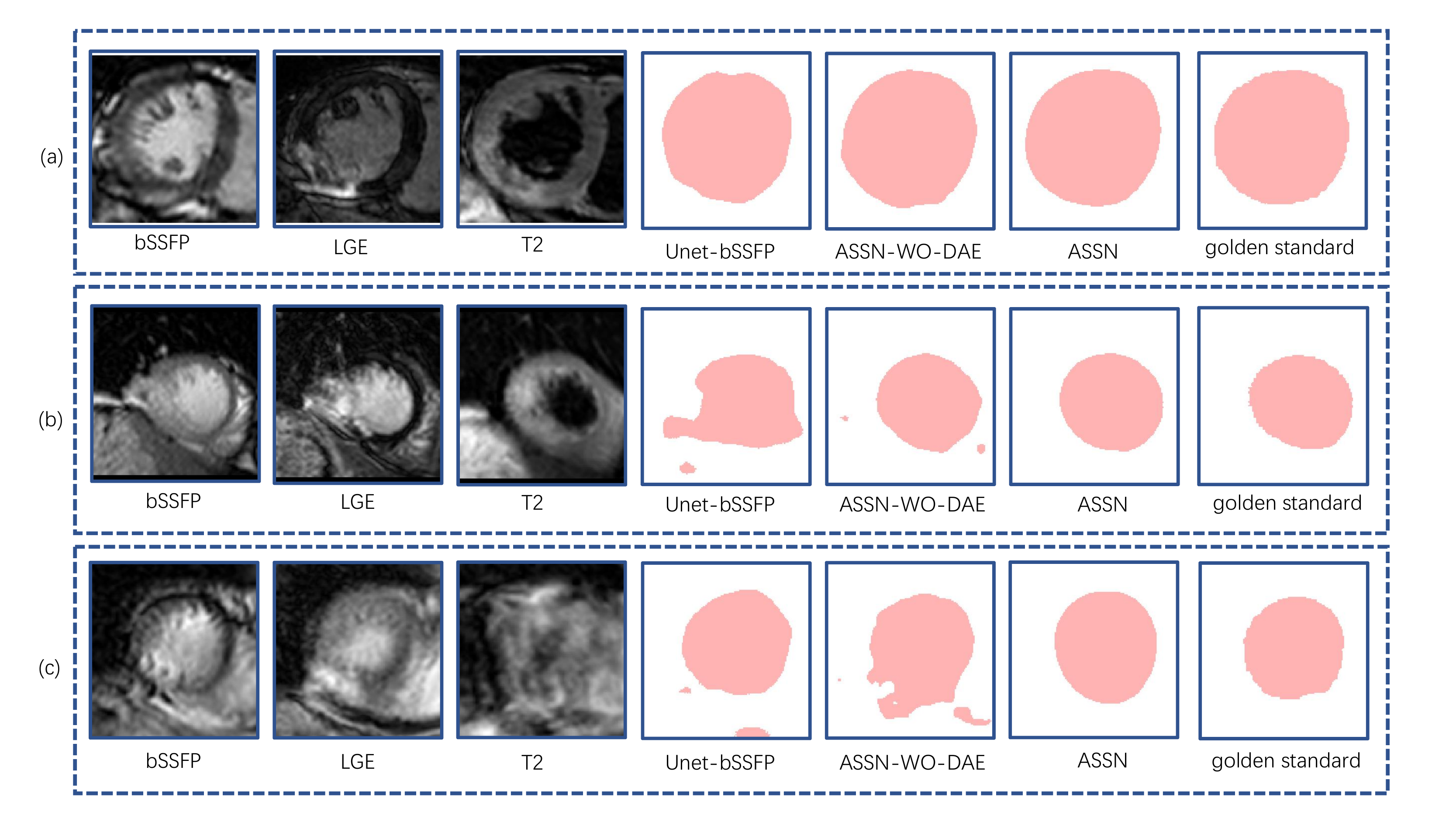}
		\caption{Comparison of the proposed method and other baseline methods. Image of (a) is a normal case, where all three methods can achieve a reasonable segmentation result. However, the image of (b) and (c) are challenging cases, where both Unet-bSSFP and ASSN -WO-DAE failed to generate realistic results but the proposed ASSN method showed good robustness.}
		\label{fig:4}    
	\end{figure}

	To evaluate the performance of the ASSN, the Dice score and Hausdorff distance between the predicted label and gold standard label were calculated. Table \ref{tab:1} shows the performance of three different methods: 
	\begin{itemize}
	    \item {Unet-bSSFP}: The Unet which is trained by using bSSFP images. We implemented this method because the bSSFP can provide a relatively clear boundary of the LV. 
	   \item {ASSN-WO-DAE}: Our ASSN network but without DAE. 
	   \item  ASSN: Our proposed anatomical structure segmentation network. 
	\end{itemize}
 Compared to Unet-bSSFP, the methods (ASSN-WO-DAE and ASSN) using multi-modality CMRs can achieve better performance in both terms of Dice score and Hausdorff distance. Additionally, although the ASSN-WO-DAE obtained comparable result to ASSN in term of the Dice score, the ASSN still achieved almost $5$ mm improvement in the Hausdorff distance. Moreover, Figure \ref{fig:4} presents a series of visual results. One can see the results of  Unet-bSSFP and ASSN-WO-DAE were easily affected by the quality of CMRs, while our ASSN can generate results with plausible shape. This demonstrates the benefit of integrating DAE to the segmentation network.
	\begin{table*}[tb] 
			\capbtabbox[0.9\textwidth]{
				\begin{tabular}{ccc}
					\hline
					Method &  Dice (\%) & Hausdorff (mm)\\
					\hline
					Unet-bSSFP  & 93.77	 $\pm$ 2.55   &  10.72	 $\pm$ 9.02  \\
					ASSN-WO-DAE  & 	96.21 $\pm$ 3.28 &   8.25 $\pm$ 9.06 \\
					ASSN  & 	\textbf{96.99 $\pm$ 1.49} 	 &  \textbf{3.04	 $\pm$ 0.76} \\
		
					\hline
				\end{tabular}
			}{
				\caption{Dice score and Hausdorff of the proposed method and other baseline methods on the validation set.}
				\label{tab:1}
			}
		
	\end{table*}
	
	\subsubsection{PRSN:}
\begin{figure}[hbt] 
		\centering
		\includegraphics[width=0.9\textwidth]{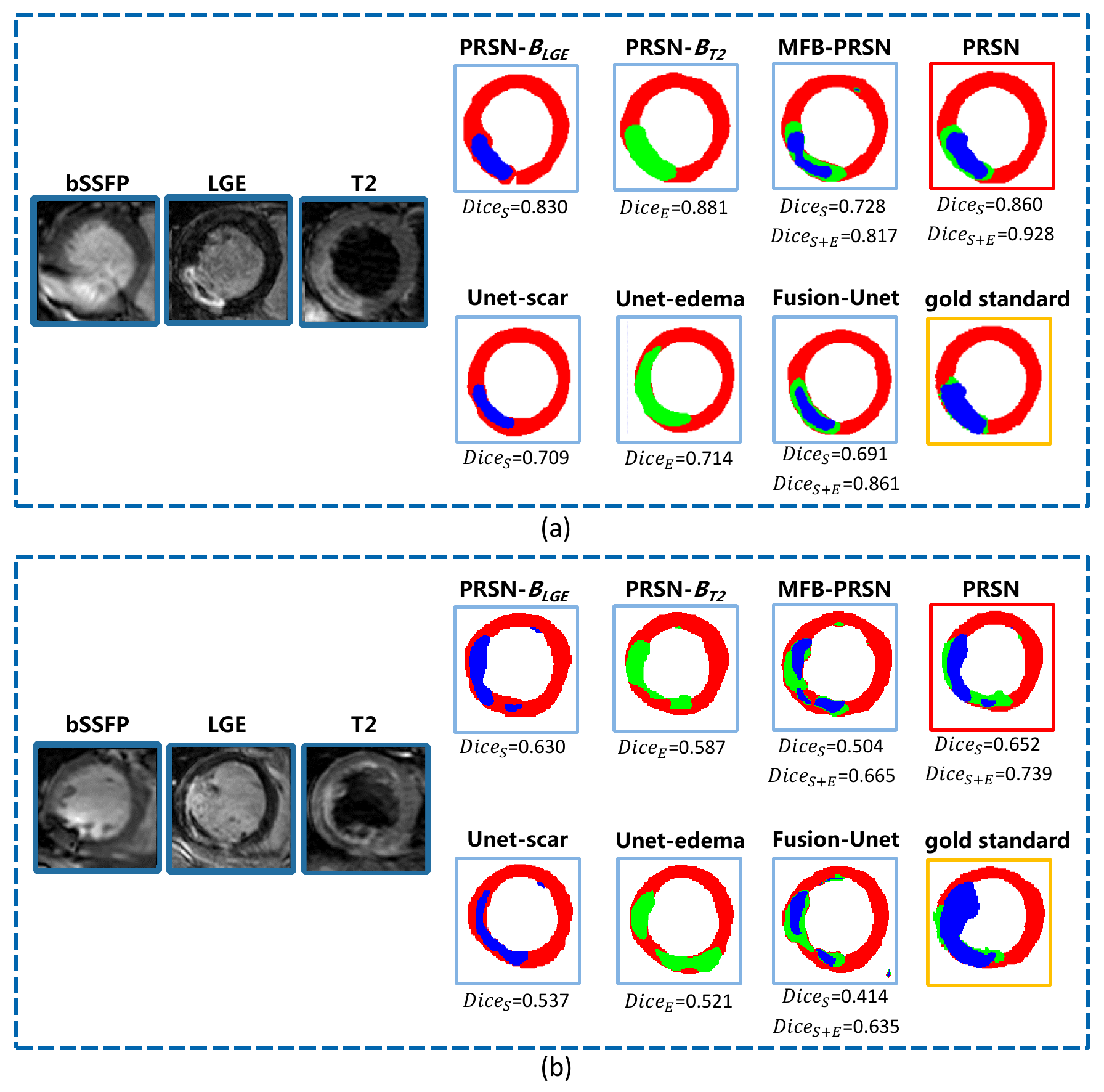}
		\caption{Visualization of the pathology segmentation results. The $Dice_S$, $Dice_E$  and $Dice_{S+E}$ refer to the Dice score of predicted scar (blue), edema (green) and scar $+$ edema region, respectively. Image of (a) is an easy case, while the image of (b) is a more challenging one. (The reader is referred to the colourful web version of this article.)}
		\label{fig:5}    
	\end{figure}

	The performance of the PRSN is evaluated by the Dice score of scar and scar $+$  edema region. Table \ref{tab:2} shows seven different segmentation methods based on our extracted LV. 
    \begin{itemize}
        \item {Unet-scar}: Unet which is trained on $C_{LGE}$ datas for scar segmentation.
        \item {Unet-edema}: Unet which is trained on $C_{T2}$ datas for edema segmentation.
        \item {PRSN-$B_{T2}$}: The $B_{T2}$ branch of PRSN.
        \item {PRSN-$B_{LGE}$}: The $B_{LGE}$ branch of PRSN.
        \item  {Fusion-Unet}: Unet which is implemented by using input-level fusion strategy \cite{zhou2019review}.
        \item  {MFB-PRSN}: PRSN which is implemented by using MFB (summation-product-maximization) fusion strategy in $B_{Main}$ \cite{zhou2020hi}.
        \item  PRSN: Our proposed pathological region segmentation network.
        
    \end{itemize}
	Among these methods, the Unet-scar, Unet-edema, PRSN-$B_{T2}$ and PRSN-$B_{LGE}$ can be considered as the mono-modality methods, while the MFB-PRSN, Fusion-Unet and PRSN are multi-modality methods. Overall, the multi-modality methods achieved better results than the mono-modality methods in scar segmentation. This reveals the advantage of using multi-modality images for pathology segmentation. Meanwhile, compared to MFB-PRSN which uses the summation-product-maximization fusion strategy for feature fusion, our PRSN achieved almost 4\% and 3\%  improvement in scar and scar $+$  edema region, respectively. This indicates the advantage of our proposed CAFB. In addition, Figure \ref{fig:5} demonstrates visual results of different methods.
	
	\begin{table*}[tb] 
		\capbtabbox[0.9\textwidth]{
			\begin{tabular}{ccc}
				\hline
				Method &  scar (\%) &  edema (\%)\\
			\hline
			Unet-edema & 	N/A & 61.42	 $\pm$ 11.86   \\
			Unet-scar & 56.38 $\pm$ 23.36   &  N/A  \\
		
			PRSN-$B_{T2}$  & N/A   & 64.37	 $\pm$ 11.25  \\
			PRSN-$B_{LGE}$  &55.65	 $\pm$ 24.97   & N/A  \\
			\hline
				&  scar (\%) & scar $+$  edema (\%)\\
			\hline
			Fusion-Unet  & 57.50	 $\pm$ 23.09   &  66.61	 $\pm$ 12.79  \\
			MFB-PRSN  & 59.56	 $\pm$ 25.18   &  68.59	 $\pm$ 12.33  \\

			PRSN & 	\textbf{64.09 $\pm$ 25.96} 	 &  \textbf{70.24 $\pm$ 12.98} \\
			
			\hline
			\end{tabular}
		}{
			\caption{Dice scores of the proposed method and other baseline methods on the testing set. N/A indicates the segmentation result was not provided}
			\label{tab:2}
		}
		
	\end{table*}
	
	\section{Conclusion}
	In this work, we proposed a cascade multi-modality pathology segmentation framework. It has been evaluated on scar and edema segmentation of CMR images. The experimental results show our CAFB is capable in fusing complementary information from multi-sequence CMRs to boost the pathology segmentation performance. Besides, we present the advantage of using DAE to reconstruct the segmentation result with plausible shape. Future research aims to investigate the performance of the framework on other pathological datasets. 

\bibliographystyle{splncs04}

\bibliography{ref}

\end{document}